\documentclass[aps,prb, twocolumn]{revtex4-1}
\usepackage{amsmath}
\usepackage{amsfonts}
\usepackage{amssymb}
\usepackage{dcolumn}
\usepackage{bm}
\usepackage{graphicx}
\usepackage{adjustbox}
\usepackage{epstopdf}
\usepackage{color}
\usepackage{xcolor}
\usepackage{textcomp}
\usepackage{gensymb}
\usepackage{adjustbox}
\usepackage{ulem}
\usepackage{comment}
\definecolor{deep-blue}{rgb}{0.17, 0.17, 0.89}

\def\BaOs  {Ba$_2$NaOsO$_6$}
  \def \Na {$^{23}$Na }

   \def \ie {{\it i.e.} }

  \def \etal {{\it et al.}}

\begin{document}
\title{First Principles calculations of the EFG tensors of Ba$_2$NaOsO$_6$, a Mott insulator with strong spin orbit coupling}
\author{Rong Cong$^{1}$, Ravindra Nanguneri$^{2}$, Brenda Rubenstein$^{2, \dag}$, and V. F. Mitrovi\'c$^{1, \dag}$}
\address{$^{1}$Department of Physics, Brown University, Providence, Rhode Island 02912, USA}
\address{$^{2}$Department of Chemistry, Brown University, Providence, Rhode Island 02912, USA}

\date{\today}
\begin{abstract}
We present first principles calculations of the electrostatic  properties of Ba$_2$NaOsO$_6$ (BNOO), a 5$d^1$ Mott insulator with strong spin orbit coupling (SOC) in its low temperature quantum phases. 
In light of recent NMR experiments showing that BNOO develops a local octahedral distortion that is accompanied by the emergence of an electric field gradient (EFG) and precedes the formation of long range magnetic order {[Lu  \etal, Nature Comm.   {\bf 8}, 14407 (2017), Liu  \etal,  Phys. Rev. B {\bf 97}, 224103 (2018), Liu  \etal, Physica B {\bf 536}, 863 (2018)]}, we calculated BNOO's EFG  tensor for  several  different model distortions.  The local orthorhombic distortion that we identified as mostly strongly agreeing with experiment  corresponds  to a Q2 distortion mode of  the  Na-O octahedra, in agreement with conclusions  given in {[Liu  \etal,  Phys. Rev. B {\bf 97}, 224103 (2018)]}.  Furthermore, we found that the EFG  is insensitive to the type of underlying magnetic order. By combining NMR results with first principles modeling, we have thus forged a more complete understanding of BNOO's structural and magnetic properties, which could not be achieved based upon experiment or theory alone.  
\end{abstract}

\maketitle

\section{Introduction}
\label{sec-introduction}
  \vspace*{-0.20cm}
Magnetic Mott insulators with strong spin orbit coupling \cite{Kim08Nov,J_eff_half_MI_2,Jeff_half_iridates,5d_IMT_CMS_RP_series,weyl} have long been a prime focus of strongly correlated materials research because of the complex interplay among their spin, orbital, and charge degrees of freedom. While 3$d$ systems have been studied intensively,\cite{RevModPhys.70.1039} much less is known about 4$d$ and 5$d$ systems, in which the more delocalized $d$ electrons, weaker correlations, and larger spin orbit coupling (SOC) effects
compete to give rise to rich magnetic and electronic phases, including multipolar magnetic ordering, SOC assisted Mott insulators, and topological insulators  \cite{Chen_PRB_2010,trivedi_2017,SOC_U_physics_2016,balents_SOC_review_2014,TI}.

In 5$d^1$ transition metal oxides with strong spin orbit coupling, the lower energy t$_{2g}$ triplet can be regarded as a pseudospin operator  $L_{\rm eff}$ = -1, which gives rise to the ground state $J_{\rm eff}=\frac{3}{2}$ quartet together with $S = \frac{1}{2}$  \cite{goodenough1968spin, Kim08Nov}.  Examples of such oxides include
the magnetic insulating double perovskite osmium compounds Ba$_2$NaOsO$_6$ (BNOO) and Ba$_2$LiOsO$_6$, which have  similar structural and electronic features, but very different magnetic properties.

The ground state of 5$d^1$ Ba$_2$LiOsO$_6$ is antiferromagnetic with an effective magnetic moment of $\mu_{\rm eff}\approx 0.7  \mu_B$,  which is  much smaller than the spin only value, indicating the presence of strong SOC \cite{Balents_2017}. Muon spin relaxation measurements also reveal a spin-flip transition in the applied magnetic field in the vicinity of  5.5 T at 2 K \cite{steele2011low}. Although Ba$_2$LiOsO$_6$'s analog, BNOO,  is also a Mott insulator with an on site Coulomb repulsion $U$ of \mbox{$\approx 3.3$ eV} and inter-site hopping $t$ of \mbox{$\approx 0.05$ eV}, \cite{erickson2007FM} what distinguishes BNOO is its substantial net ferromagnetic moment of 0.2 $\mu_B$ per osmium atom  \cite{NPJ_Q_M}   and negative Curie-Weiss temperature \cite{erickson2007FM,stitzer2002crystal}. Moreover, the [110] easy axis of the ferromagnetic ground state of BNOO   is also unique since standard Landau theory can only give rise to [100] or [111] easy axes when cubic symmetry is present \cite{Chen_PRB_2010}.

As has recently been illuminated via nuclear magnetic resonance (NMR) experiments, Ba$_2$NaOsO$_{6}$ undergoes multiple phase transitions to states that exhibit exotic order with decreasing temperature $(T)$\cite{Lu_NatureComm_2017,Liu_Physica_2018}. At high temperatures, Ba$_2$NaOsO$_{6}$ is a paramagnet (PM) with perfect {\it fcc} cubic symmetry and no oxygen octahedral distortion, as sometimes occurs in other transition metal oxides. Cubic symmetry, along with the lattice constant and the Oxygen Wyckoff position, uniquely fixes the high temperature, undistorted structure. Upon lowering the temperature (e.g., below 13 K at 15 T), local octahedral distortion, identified as broken local point symmetry (BLPS) \cite{Lu_NatureComm_2017}, onsets while the global symmetry of the unit cell still remains cubic. 
This BLPS phase  precedes the formation of long-range magnetic order.
At even lower temperatures, below $\sim10 \, {\rm K}$,  local orthorhombic octahedral distortion is found to coexist 
with two-sublattice canted ferromagnetic (cFM) order   \cite{Lu_NatureComm_2017, Liu_PRB_2018}. This transition into the magnetically ordered state  is thought to be a tetrahedral to orthorhombic transition. One hypothesis is that formation of orbital order, caused by purely electronic Coulomb interactions within the Os 5$d$ orbitals, 
drives this structural, Jahn-Teller type transition \cite{KK_1,KK_2}.

To identify  the exact  local structural distortion pattern, in our previous work in  \mbox{Ref. [\onlinecite{Liu_PRB_2018}]}, we utilized a point charge approximation combined with six different structural distortions (Models A-F in the following, see Fig.~\ref{visina8}) to simulate the EFG tensor.  However, this approach does not allow us to distinguish  between whether displacements of the actual ions or distortions of the ion charge density  are responsible for  the appearance   of the   finite EFG at the Na site  {\cite{Liu_PRB_2018, Goncalves12}}. Moreover, this method neglects crucial physics when approximating the EFG, which calls conclusions derived from it into question. First, using the point charge approximation, it is not clear how to assign charge to the different ions. Second, calculations of electric potentials within finite boxes are never fully converged. Third, the formation of bonds will also influence the local potential. Lastly, Ba$_2$NaOsO$_6$ is a Mott insulator with strong spin orbit coupling and the interplay between spin and charge may also influence the local potential.

To remedy the aforementioned problems, in this paper, we report a DFT+U and hybrid DFT study of the $5d^1$ strongly spin orbit coupled transition metal oxide Ba$_2$NaOsO$_6$ in its double perovskite   structure, highlighting how computational  
 methods can be combined with  NMR  data to elucidate the structural distortion and/or charge density  patterns in the low temperature phase of this material. 
This direct consideration of the observed structural  distortions  distinguishes our work from all   previous first principles calculations \cite{Pickett_2007, Pickett_2015,Pickett_2016}. 
We find that, within our DFT calculations, the orthohombic local distortions embodied by models characterized by the dominant displacement of oxygen ions along the cubic axes of the perovskite reference unit cell \cite{Liu_PRB_2018} (Models A, B, and F2) for certain distortion values are possible candidates for the BLPS phase, with Model A best matching the NMR data, consistent with findings in \mbox{Ref. [\onlinecite{Liu_PRB_2018}]}.

This paper is organized as follows. The theory of NMR that relates our DFT calculations to NMR observables is first described in Section~\ref{sec:NMRtheory}. In Section~\ref{sec:CompApproach}, we then detail how we performed our first principles simulations. In Section~\ref{Results}, we present our numerical results for the EFG tensors produced by the six different BLPS distortion models.   Lastly, in Section~\ref{Conc}, we conclude with a summary of our current findings and their bearing on the physics of related materials. 

\section{Deducing the electric field gradient tensor from NMR}
\label{sec:NMRtheory}

The orthorhombic distortion engenders a single, structurally equivalent Na NMR site \cite{Liu_PRB_2018}. What is known about this distortion is that it creates an electric field gradient (EFG) due to the charge density, as well as magnetic transfer hyperfine fields due to exchange coupling, at the \Na sites. Without a distortion, the   EFG tensor at the \Na site would be traceless and zero for a cubic structure. The Na nuclei possess a finite quadrupole moment, owing to their 
large nuclear spin ($I=\frac{3}{2}$). The quadrupolar electric potential of the charge distortion couples to the nuclear quadrupole to split the nuclear spin multiplet.   {Moreover, it is a significant advantage that the  \Na sites are away from the magnetic moment carrying Os sites. This is because the super-exchange   between Os ions is mediated via oxygen orbitals and the transfer-hyperfine interaction between Na nuclei and magnetic moments allows one to see such an effect. \cite{Lu_NatureComm_2017}.}  
NMR  at the \Na sites can thus be a sensitive local probe of the charge distortion and the magnetic ordering at the osmium sites.

As alluded to above, the NMR spectrum 
of asymmetric nuclei with finite quadrupole moments undergoes a  splitting when there is a non-zero electric field gradient (EFG) at the nuclear site being studied.\cite{Liu_PRB_2018} The EFG is formally characterized by the EFG tensor, $\nabla\bf{E}$, which is a symmetric ($\nabla\times\bf{E} =0$) and traceless ($\nabla \dot \bf{E} =0$) rank-2 tensor.\cite{volkoff1952nuclear} All of the information present in an EFG tensor is contained within its eigenvalues, $V_{xx}$, $V_{yy}$, and $V_{zz}$, and their corresponding eigenvectors after diagonalization. The eigenvalues are named according to the common convention  $|V_{zz}|>|V_{yy}|>|V_{xx}|$. The asymmetry  parameter $\eta$ is defined as $\eta$ = $(V_{xx}-V_{yy})/V_{zz}$ such that   $0<\eta<1$, by definition. The eigenvectors, also called the principal axes, define a right-handed, rectangular EFG coordinate system $O_{XYZ}$ that does not necessarily align with that defined by the  crystalline axes $(a,b,c)$, $O_{abc}$. We note that, since the diagonalization of a rank 2 tensor only determines the orientation of the principal axes but not their relative signs, the positive direction of the corresponding EFG coordinate axes remains undetermined. 
This uncertainty, however, may be eliminated by comparing to experimental results. 

In NMR spectra, the splitting $(\delta_q)$, corresponding to the frequency difference between adjacent quadrupole satellite transitions,   can be  written in terms of the EFG tensor parameters  as 
\begin{equation}
    \delta_q = \left  | \frac{1}{2}\nu_Q (3 \cos^{2} \theta - 1 + \eta \sin^2\theta \cos2\phi) \right |,
\end{equation}
where $\nu_Q$ = $eQV_{zz}/(2h)$ with $eQ$ the nuclear quadrupole moment and $h$ Planck's constant. The angle $\theta$ is the angle  between the applied field $H$ and $V_{zz}$,  and $\phi$ is the standard azimuthal angle of a spherical coordinate system defined by $O_{{XYZ}}$.
As 
 only the magnitude of $\nu_{Q}$ affects the triplet splitting in NMR experiments (see Appendix A of Ref. \onlinecite{Liu_PRB_2018}),  in this work  we treat both positive and negative $\nu_{Q}$ values as identical.  

\begin{center}	
 %
\begin{figure}[t]
  \vspace*{-0.0cm}
\begin{minipage}{1\hsize}
 \centerline{\includegraphics[scale=0.48]{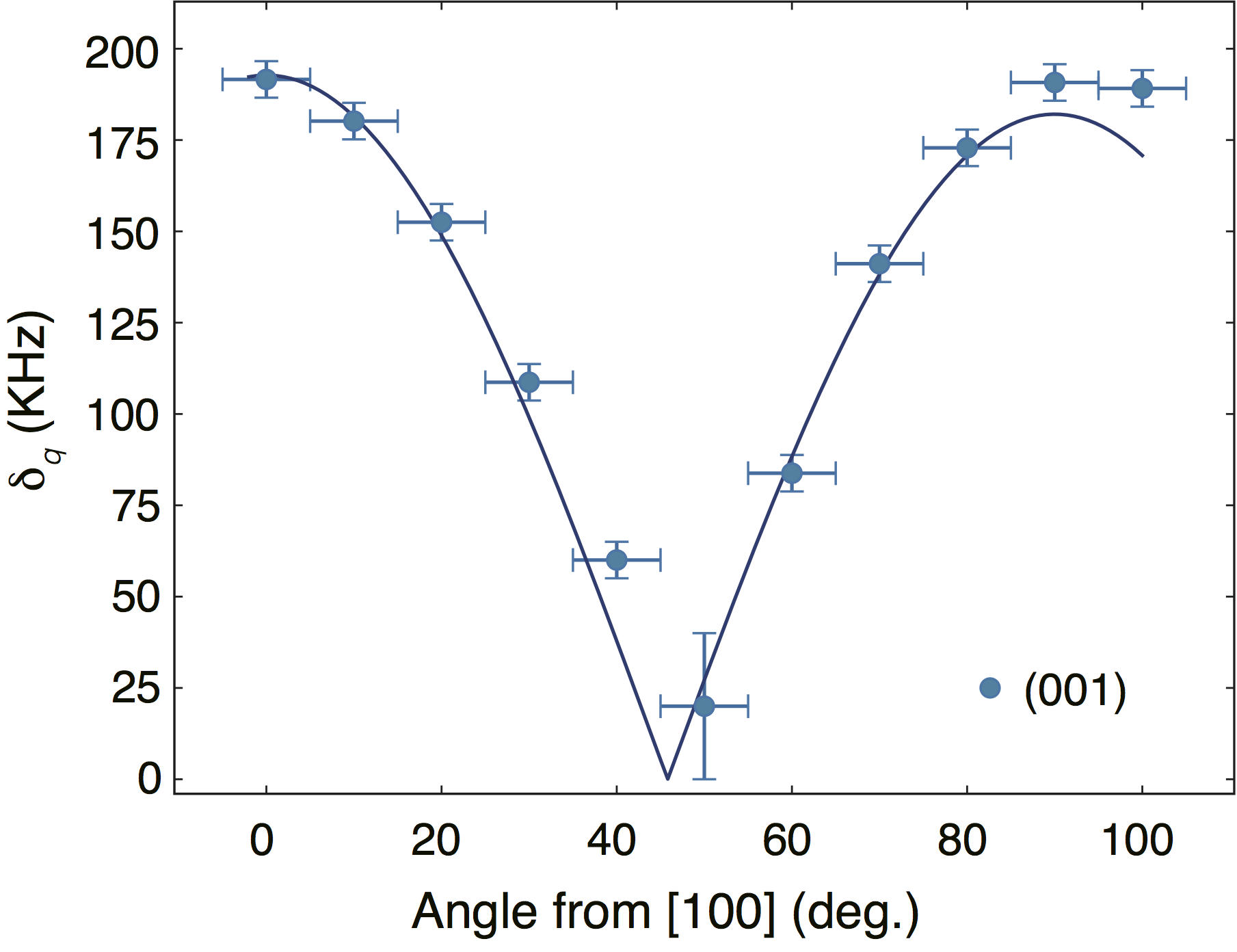}} 
\begin{minipage}{1\hsize}
 \vspace*{-0.1cm}
\caption[]{\label{fig:rotation} \small 
The mean peak-to-peak splitting $(\delta_{q})$ between any two adjacent peaks of the quadrupole split Na spectra in the BLPS phase as a function of the angle between the [100] crystal axis and the applied magnetic field $(H)$. The blue dots denote the measured angular dependence of the splitting when the sample is rotated in the (001) plane in  a  4.5 T applied field at 5 K. The solid line is the calculated angular dependence 
  using EFG parameters obtained from Model A, Case 3 (Model A.3), as described in the text.  }
 \vspace*{-0.0cm}
\end{minipage}
\end{minipage}
\end{figure}
%
\end{center}

Following a similar analysis to that detailed in Ref.~\onlinecite{Liu_PRB_2018}, in which different possibilities for the relative alignment between the 
coordinate systems defined by the 
crystalline axes $O_{{abc}}$ and those of the EFG $O_{{XYZ}}$ were considered, we obtained two possibilities for the EFG parameters ($\eta$, $\nu_Q$, and the principal axes) consistent with our observations. For the first possibility, the principal axes of the EFG tensor are $V_{zz} || c$, $V_{yy} || a$, and $V_{xx} || b$, $\nu_Q\approx\pm199 \, {\rm kHz}$, and $\eta\approx0.88$. For the second possibility, $V_{zz} || a$, $V_{yy} || c$, and $V_{xx} || b$,  $\nu_Q\approx\pm192 \, {\rm kHz}$, and $\eta\approx1$. Both sets of these EFG parameters successfully reproduce the experimental rotation pattern in the (001) plane depicted in Fig.~\ref{fig:rotation}, as well as  those in the (010) and $(1 \bar{1} 0)$ planes, described in  detail in \mbox{Ref. [\onlinecite{Liu_PRB_2018}]}. In  a material with global cubic symmetry, the crystalline axes can be   distinguished as a result of a weak symmetry-breaking field that favors one direction over the others. 
 We deduce that a   source of such a symmetry breaking field is provided by  the strain from the way the sample was mounted on the flat platform, which was always parallel to the specific face of the crystal. 
Given the similar $\eta$ and $\nu_Q$ values we obtained for these two cases, we can conclude that NMR experiments\cite{Lu_NatureComm_2017,Liu_PRB_2018} have shown that the principal axis, $V_{zz}$, of the EFG tensor either aligns with the $a$ or $c$ crystalline axes with $\eta$ close to 1 and $\nu_Q \approx\pm 190 - 200 \, {\rm kHz}$, and that the principal axes of the EFG must align with the cubic axes of the perovskite reference unit cell. 
We therefore compare these parameters with the average values of the calculated EFG parameters in the following sections. 

 \section{Computational Approach}
\label{sec:CompApproach}
   \vspace*{-0.20cm}
  
To perform the calculations that follow, we used the   Vienna Ab initio Simulation Package (VASP),  complex version 5.4.1/.4, plane-wave basis DFT code \cite{vasp_1,vasp_2,vasp_3,vasp_4}. The exchange-correlation functionals employed were the Generalized-Gradient Approximation PW91 \cite{GGA} and Perdew-Burke-Ernzerhof (PBE) \cite{PBE} functionals, both supplemented with two-component spin orbit coupling. We used $500$ eV as the plane wave basis cutoff energy and we sampled the Brillouin zone using an $8\times 8\times 8$ k-point grid. The criterion for stopping the DFT self-consistency cycle is a $10^{-5}$ eV difference between successive total energies. To facilitate convergence of the k-space charge density, we smooth our Fermi functions by allowing fractional occupations of frontier orbitals in our self-consistent calculations using the Methfessel-Paxton (MP) smearing technique \cite{MP_smearing}.

   \onecolumngrid
 \begin{center}	
 %
\begin{figure}[h]
  \vspace*{-0.0cm}
\begin{minipage}{0.98\hsize}
 \centerline{\includegraphics[scale=0.47]{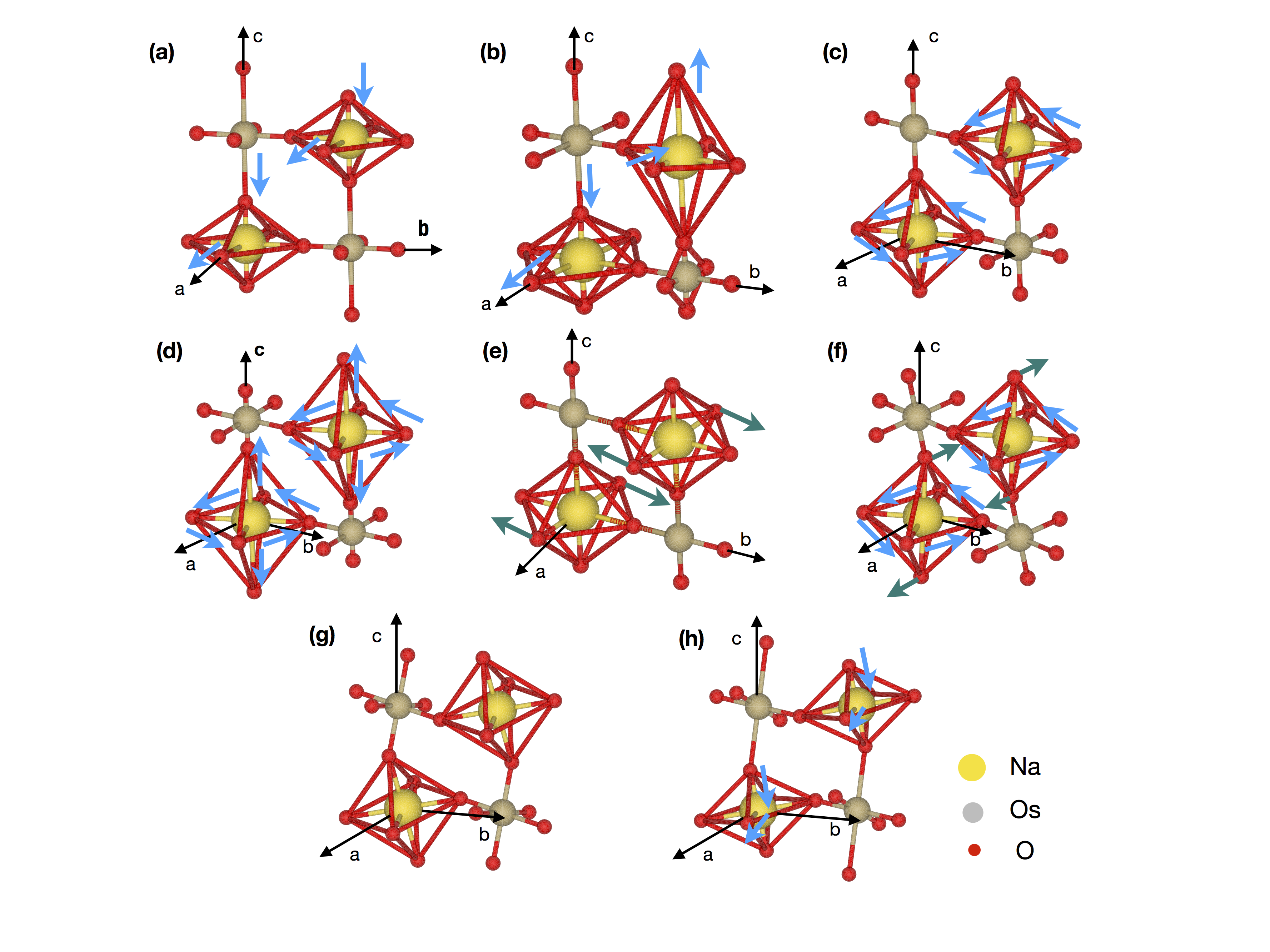}} 
\begin{minipage}{.98\hsize}
 \vspace*{-0.4cm}
\caption[]{\label{visina8} \small 
Illustration of the different models of lattice distortion discussed in the text. (a) Model A: uniform compression (left) or elongation (right), (b) Model B: two-sublattice compression and elongation, (c) Model C: rotation in the $ab$ plane, (d) Model C2: elongation of Model C along the c axis, (e) Model D: tilt distortion, (f) Model E: rotation plus tilt distortion, (g) Model F: GdFeO$_3$-type distortion, and (h) Model F2: Model A type distortion applied to Model F.  Blue arrows indicate elongation, compression, or rotation, and green arrows indicate tilt distortion on a particular plane. Na atoms are depicted in yellow, Os atoms in gray, and O atoms in red.  }
 \vspace*{-0.3cm}
\end{minipage}
\end{minipage}
\end{figure}
%
\end{center}
  \vspace*{-0.10cm}
\twocolumngrid

In DFT+U calculations, two tunable parameters, $U$ and $J$, are employed. $U$ describes the screened-Coulomb density-density interaction acting on the Os 5$d$ orbitals and $J$ is the Hund's interaction that favors maximizing $S^z_{total}$ \cite{hund}. 
In all of the calculations that follow, we set $U=3.3$ eV and $J=0.5$ eV based upon measurements from Ref. \onlinecite{erickson2007FM} and then tested that the calculated EFG parameters are insensitive to the precise values of $3.3  < U < 5.0$ eV for fixed $J = 0.5$ eV and $0.5 < J < 1.0$ eV for fixed $U = 3.3$ eV. As demonstrated in \mbox{Appendix \ref{CheckDFT}}, we find that our EFG results are largely invariant over this wide range of $U$ and $J$ values, justifying our use of the DFT+U method for studying this problem.

In order to increase the computational efficiency of our simulations, we employed projector augmented wave (PAW)\cite{PAW_Blochl,PAW_vasp} pseudopotentials (PPs) in both our DFT+U and hybrid functional calculations. We tested six different types of PPs labeled
PP1: Ba$_{sv}$+Na+Os+O, PP2: Ba$_{ sv}$+Na$_{pv}$+Os$_{pv}$+O$_{s}$, PP3: Ba$_{sv}$+Na$_{pv}$+Os+O, PP4: Ba$_{ sv}$+Na+Os$_{pv}$+O, PP5: Ba$_{ sv}$+Na+Os+O$_{s}$, and PP6: Ba$_{ sv}$+Na+Os$_{pv}$+O$_{s}$. The subscripts $pv$ and $sv$ indicate that $p$ and $s$ semi-core orbitals are also included in the valence electron set, and $s$ indicates that the PP is softer than the standard version. It is generally expected that including more valence electrons explicitly will give rise to more accurate results. For EFG calculations, this implies that high quality PAW basis sets are typically required, which indicates that semi-core electrons are important. Indeed, we found that explicitly including $p$ and $5d$ electrons on the Os atom, as embodied in the PP6 pseudopotential, is key to reproducing experimental EFG parameters. We therefore used this pseudopotential throughout this work.

BNOO belongs to the space-group $Fd\bar{3}m$ and has an {\it fcc} primitive cell. When considering distortions which give rise to two magnetically distinct Na sites with cFM order, we must use a more complicated cubic unit cell with lattice constant \mbox{8.287 \AA}  comprised of four primitive cells in order to calculate the EFG tensor. The unit cell and atom positions are depicted in the \mbox{Appendix \ref{Unitcell}}. 

In addition to the DFT+U method, we also used the 
PBE0\cite{PBE0}
functional to calculate the EFG tensor of representative BLPS structures.
For PBE0, $25\%$ 
of the DFT exchange energy is replaced with the exact Hartree-Fock (HF) exchange energy, which is expected to better capture the long-range behavior of the exchange potential $v_{x}$ and to provide a check on our DFT+U results. For computational expediency, a single {\it fcc} primitive cell with an FM Os moment was used in all of the more expensive hybrid calculations. We note that, while hybrid DFT may serve as a check on DFT+U results, hybrid DFT is also not a purely \textit{ab initio} method due to the percentage of HF exchange energy that has to be tuned in the functional.

  \vspace*{-0.40cm}

The general outline for the calculations we performed is described in the following. 
 We first carried out single self-consistent or `static' calculations with GGA+SOC+U with fixed structures for the models labelled A-F2 (see Fig.~\ref{visina8}), most of which were previously considered within the point-charge approximation in Ref. \onlinecite{Liu_PRB_2018}. In these calculations, the magnitude of the distortion is varied by hand. Non-collinear, cFM initial magnetic moments are imposed for the two osmium sublattices in the directions determined in Ref. \onlinecite{Lu_NatureComm_2017}. In most cases, the converged magnetic moments (orbital plus spin) continue to point along the cFM directions. For certain models, namely A, B, and F2, we obtain EFG parameters similar to those observed in experiment. For these, we perform tests with SOC and U to determine their effects. We also performed hybrid DFT calculations to check the robustness of the DFT+U results for these models,  with the exception of Model B, which cannot be realized in the {\it fcc} primitive cell.

The EFG tensor in our DFT+U calculation is given by the gradient of the electric fields, or the second partial derivatives of the scalar potential at the Na nuclear site. It is obtained from the DFT charge density as a post-processing step by solving Poisson's equation for the scalar potential.  
  \vspace*{-0.40cm}
\section{Results and Discussions}
\label{Results}
  \vspace*{-0.10cm}
 
We performed our DFT+U calculations on all of the model distorted structures depicted in Fig.~\ref{visina8} that were proposed in Ref.~\onlinecite{Liu_PRB_2018}. These distorted structures include: {\bf I.} Identical orthorhombic distortions on both Na sites in the unit cell such that both Na sites remain structurally equivalent (A);  {\bf II.}  Different and opposite orthorhombic distortions on the two Na sites leading to structurally inequivalent Na sites (B);  {\bf III.}  Pure rotational distortion in the $(a,b)$-plane which keeps the Os-O and Na-O bond lengths unchanged, but deforms the Os-O-Os angles (C);  {\bf IV.}  C type distortion plus elongation of the Os-O bond along the $z$ axis (C2);  {\bf V.} Tilt distortion of the Os-O bond of the $\alpha$ axis on the $\beta$-plane. $\alpha$ can represent $a$, $b$, or $c$, while $\beta$ can represent ($a$,$b$), ($a$,$c$), or ($b$,$c$) (D);   {\bf VI.} Rotational distortion in the ($a$,$b$)-plane and tilt distortion in the ($a$,$c$)-plane (E);  {\bf VII.} GdFeO$_3$ type \cite{glazer1972classification} 
distortion with rigid octahedra (F); and   GdFeO$_3$ type distortion with flexible octahedra (F2).

\subsection{Static Calculations}
\label{subsec:static}
Given one of the structures that follow, we calculate its EFG using VASP. 
Results are shown in  the tables below. The meaning of the calculated  quantities is consistent with the definitions established in \mbox{Sec. \ref{sec:NMRtheory}}. That is, 
for all of the tables in this section, $V_{zz}$ indicates the direction in the coordinate system   $O_{{XYZ}}$  that  aligns with the principal EFG axis corresponding to the leading eigenvalue,  $\eta$ represents the asymmetry factor, and $\nu_Q$ represents the electric quadrupolar splitting parameter, \ie, the maximum frequency difference between adjacent quadrupole satellite transitions. 
We define the distortion as a percentage relative to the Na-O distance, \mbox{2.274 \AA}, of the undistorted bond. Here, we also define the   elongation deformation as   positive   and the compression deformation as negative.

\subparagraph{{\bf Undistorted Case}}
In this case, the structure is given by the experimentally-determined, high temperature cubic structure without any distortion. The $U$ value of 3.3 eV is taken from Ref.~\onlinecite{erickson2007FM}. We found that, without SOC, there is no splitting ($\nu_Q=0$) and $V_{zz}$, $V_{yy}$, and $V_{xx}$ are all zero, so that $\eta$ and the principal axis are undetermined. This is consistent with the fact that, for a perfect cubic structure, the EFG is zero. Including SOC and imposing the two sub-lattice cFM order described in Ref.~\onlinecite{Lu_NatureComm_2017} results in EFG parameters that are no longer zero, but far too small to account for the desired 190-200 kHz splitting seen in our NMR experiments.

\begin{table}[h]
\begin{tabular}{|c|c|c|c|}
\hline
\textbf{Method} & \boldmath{$V_{zz}$} & \boldmath{$\eta$} &\boldmath{$\nu_Q \, {\rm (kHz)}$}           \\ \hline
GGA & n/a & n/a & 0 \\ \hline
GGA+U & n/a & n/a & 0 \\ \hline
GGA+SOC & a & 0.813 & -0.5 \\ \hline
GGA+SOC+U & a & 0.302 & -25 \\ \hline
\end{tabular}
\caption{EFG parameters for the undistorted structure using different methods.}
\label{tab:1}
\end{table}

To determine the origin of the non-zero EFG parameters, we considered the models of local distortion proposed in Ref.~\onlinecite{Liu_PRB_2018}. 
 The initial magnetic moments were set as indicated in Ref.~\onlinecite{Lu_NatureComm_2017}, where the staggered moments  alternate symmetrically  about the [110] axis  from neighboring layer to neighboring layer along the $c$-axis.  
In the tables that follow, the simulation results for each of the four different Na atoms in the unit cell are given in separate rows for each set of conditions.

\subparagraph{{\bf I. Model A}}

In Model A, the Na-O bonds of the Na-O octahedra are either uniformly compressed or elongated along the original  cubic axes of the perovskite reference unit cell. 
 We found that orthorhombic distortions elongated along the $a$ axis by 0.53\% to 0.55\% and compressed along the $c$ axis by the same percentage while leaving the $b$ axis untouched can produce the desired EFG parameters, as shown for Models A.2 and A.3 (where the 2 and 3 denote models with different A-type distortion percentages) in Table \ref{tab:2} and Figure \ref{fig:rotation}.
The four Na sites have slightly different values of $\eta$ and $\nu_Q$. The difference in splitting is about \mbox{20 kHz}, which is smaller than the linewidth of the individual satellite transition in the  \Na triplet  NMR spectrum,\cite{Lu_NatureComm_2017} indicating that the broadening of the NMR spectra lines in the triplet can be attributed to slight differences in the electric field gradient at the four Na sites. Nevertheless, the assumption made earlier that the four Na atoms share the same EFG parameters still holds given the small magnitude of this difference and the fact that the NMR spectrum only reflects averages over the sites. 
 
\begin{table}[h]
\begin{tabular}{|c|c|c|c|c|c|c|}
\hline
\textbf{Case No.} & \boldmath{$\delta_a$} & \boldmath{$\delta_b$} & \boldmath{$\delta_c$}  & \boldmath{$V_{zz}$} & \boldmath{$\eta$} & \boldmath{$\nu_Q \, {\rm (kHz)}$}           \\ \hline
1 & -0.60\% & 0\% & 0.60\% & -a & 0.984 & 213 \\ 
& &  &   & -a & 0.981 & 213 \\ 
& &  &   & -a & 0.828 & 230 \\ 
& & &    & -a & 0.830 & 230 \\ \hline
\textbf{2} & \textbf{-0.54\%} & \textbf{0\%} & \textbf{0.55\%} & \textbf{c} & \textbf{1} & \textbf{-190} \\ 
& &  &   & \textbf{c} & \textbf{0.991} & \textbf{-190.5} \\ 
& &  &    & \textbf{a} & \textbf{0.818} & \textbf{209.5} \\ 
& & &    & \textbf{a} & \textbf{0.813} & \textbf{209.5} \\ \hline
\textbf{3} & \textbf{-0.525\%} & \textbf{0\%} & \textbf{0.52\%} & \textbf{a} & \textbf{ 0.981} & \textbf{183} \\ 
& & &  & \textbf{a} & \textbf{0.991} & \textbf{183} \\ 
& &    &  & \textbf{a} & \textbf{0.795} & \textbf{202} \\ 
&   &  &  & \textbf{a} & \textbf{0.790} & \textbf{203} \\ \hline
\end{tabular}
\caption{Static calculation results for the EFG parameters of Model A using GGA+SOC+U. $\delta_a$, $\delta_b$, and $\delta_c$ indicate distortions of the Na-O bond along the $a$, $b$, and $c$ axes in the crystalline coordinate system, respectively. Positive distortions indicate compression and negative distortions indicate elongation. Bold denotes those distortions with EFG parameters that best match experiment.} 
\label{tab:2}
\end{table}


\begin{table}[h]
\begin{tabular}{|c|c|c|c|}
\hline
\textbf{Method} & \boldmath{$V_{zz}$} & \boldmath{$\eta$} &\boldmath{$\nu_Q \, {\rm (kHz)}$}           \\ \hline
GGA+SOC+cFM+U & a & 0.981 & 183 \\ 
 & a & 0.991 & 183 \\ 
 & a & 0.795 & 202 \\ 
 & a & 0.790 & 203 \\ \hline
GGA+cFM+U & c & 0.950 & -211 \\ 
 & a & 0.852 & 205 \\ 
 & -a & 0.905 & 217 \\ 
 & a & 0.768 & 209 \\ \hline
GGA+U & c & 0.984 & 186 \\ 
 & c & 0.984 & 186 \\ 
& c & 0.984 & 186 \\ 
 & c & 0.984 & 186 \\ \hline
\end{tabular}
\caption{Static calculation EFG parameters for Model A with -0.525\% distortion along the $a$ axis, 0\% distortion along the $b$ axis, and 0.52\% along the $c$ axis (Model A.3) using different simulation methods. cFM stands for  the  canted ferromagnetic ordering as deduced in \mbox{Ref.  [\onlinecite{Lu_NatureComm_2017}]}.}
\label{tab:3}
\end{table}

To analyze the influence of SOC and magnetic ordering on the EFG parameters, which could not be accounted for in earlier point charge approximation calculations,\cite{Liu_PRB_2018} we calculated $V_{zz}$, $\eta$, and $\nu_{Q}$ for Model A.3 with and without SOC and non-collinear (ncl) cFM magnetization, as given in Table \ref{tab:3}. 

We found that, without SOC and cFM order, the four Na atoms have the same values of $\eta$ and $\nu_Q$. With cFM order only, the four Na atoms have different EFG parameters, all with larger splittings than observed. With SOC and cFM order, there are two distinct electronic environments as evidenced by the two sets of EFG parameters for the four Na atoms as shown in the first through fourth lines in Table \ref{tab:3}. Thus, we can conclude that the Model A local lattice distortion itself gives rise to non-zero EFG parameters at the Na sites and that the combination of cFM and SOC produce two-sublattice EFG parameters that account for the line broadening of the NMR peaks, corresponding to  the individual satellite transition within the quadrupole split   \Na triplet      \cite{Lu_NatureComm_2017}. Even though SOC and cFM order induce a distinct two-sublattice EFG tensor, its effect on the NMR observables is secondary as it only affects the line broadening and not the   quadrupole satellite line splitting. Therefore, the observed EFG parameters are overwhelmingly determined by the magnitude of the Jahn-Teller-type lattice distortion.

\subparagraph{{\bf II. Model B}} In Model B, two inequivalent Na sites emerge from two different local octahedral distortions. Based on the Model A results, we tested different ratios of local distortions. We found that, while certain cases (such as Model B.2 in Table \ref{tab:4})
produce a splitting value that matches experiment, the asymmetry factor $\eta$ does not match as well as that obtained from Model A. The best distortion found still roughly has the same ratios of distortion for the two  distinct Na  sites, which indicates that the uniform  orthorhombic  local octahedral distortion is more likely than the two-sublattice distortion to represent the BLPS phase\cite{Lu_NatureComm_2017} in BNOO.

\begin{table}[h]
\begin{tabular}{|c|c|c|c|c|c|c|}
\hline
\textbf{Case No.} &\boldmath{$\delta_a$} & \boldmath{$\delta_b$} & \boldmath{$\delta_c$} & \boldmath{$V_{zz}$} & \boldmath{$\eta$} & \boldmath{$\nu_Q \, {\rm (kHz)}$}           \\ \hline
1 & -0.53\% & 0\% & 0.55\%  & c & 0.974 & -189.5 \\
& & & & c & 0.982 & -188.5 \\
 & -0.55\%& 0\%& 0.53\%& c & 0.778 & 209.5 \\ 
& & & & a & 0.783 & 209.5 \\ \hline
2 & -0.56\% & 0\% & 0.56\%  & -a & 0.740 & 183 \\
& & & & -a & 0.725 & 183 \\
& 0.56\%& 0\%& -0.56\%& a & 0.853 & -200 \\ 
& & & & a & 0.847 & -200 \\ \hline
3 & -0.52\% & 0\% & 0.52\%  & a & 0.768 & 165.5 \\
& & & & a & 0.760 & 166.5 \\
& 0.52\%& 0\%& -0.52\% & a & 0.913 & -182.5 \\ 
& & & & a & 0.911 & -182 \\ \hline
\end{tabular}
\caption{Static calculation EFG parameters for Model B using GGA+SOC+U. The two different rows under each case condition represent the two sublattices.}
\label{tab:4}
\end{table}

\subparagraph{{\bf III. Model C}}
Model C describes the rotation of the O atoms in the ($a$,$b$)-plane. This rotation may also be accompanied by a length change in the Na-O bond along the $c$ axis, in which case we distinguish this variant as Model C2. Static calculations of Models C and C2 produce very different $\nu_Q$ and $\eta$ values for the four Na atoms. This is inconsistent with the observed quadrupolar splitting since there are only three peaks in the spectrum with a linewidth smaller than 50 kHz. 
Calculations yield a difference between $\nu_{Q}$ values that is much larger than 50 kHz, which is in striking disagreement with experiment.\cite{Lu_NatureComm_2017} Moreover, the absolute values of $\eta$ and $\nu_Q$ significantly differ from those observed experimentally. Typical examples are presented in Table \ref{tab:5}.

\begin{table}[h]
\begin{tabular}{|c|c|c|c|c|}
\hline
\boldmath{$\phi$} & \boldmath{$\delta_{c}$} & \boldmath{$V_{zz}$} & \boldmath{$\eta$} &\boldmath{$\nu_Q \, {\rm (kHz)}$}           \\ \hline
5\degree & 0\% & c & 0.970 & 87 \\ 
&  & $\approx$(a,b) dia & 0.692 & 453 \\ 
 &  & c & 0.441 & 50 \\ 
 &  & c & 0.586 & 56 \\ \hline
5\degree & 1\% & c & 0.545 & -172 \\ 
 &  & $\approx$(-a,b) dia & 0.277 & 517 \\ 
 &  & c & 0.277 & 517 \\ 
 &  & c & 0.150 & -209 \\ \hline
\end{tabular}
\caption{Static calculation EFG parameters for Models C and C2 using GGA+SOC+U. ``$\approx$ dia" means that the principal axes align more closely with the diagonal direction rather than with any crystalline axis.}
\label{tab:5}
\end{table}
 
\subparagraph{{\bf IV. Model D}}
In Model D, we considered tilt distortion in the ($a$,$c$)-plane as described in Ref.~\onlinecite{Lu_NatureComm_2017}. In addition, we also considered the tilt distortion of oxygen atoms residing on different axes tilted along the ($a$,$b$) and ($b$,$c$) planes. We found that, in all of these tilted distortion cases, the principal axes are not aligned with any of the crystalline axes. Instead, they are closer to the diagonal direction, which is labelled as ``$\approx$ dia" in Table \ref{tab:6}. Model D EFG parameters are also significantly different from those obtained via NMR. 
\begin{table}[h]
\begin{tabular}{|c|c|c|c|c|c|}
\hline
\textbf{Axis} & \textbf{Plane} & \boldmath{$\phi$} & \boldmath{$V_{zz}$} & \boldmath{$\eta$} &\boldmath{$\nu_Q \, {\rm (kHz)}$}           \\ \hline
c & (a,c) & 8.5\degree & $\approx$(a,c) ~dia & 0.789 & 359\\
 &  & & $\approx$(a,c) ~dia & 0.788 & 359\\
 &  &  & $\approx$(a,c) ~dia & 0.791 & 359\\ 
 &  & & $\approx$(a,c) ~dia & 0.790 & 359\\ \hline
c & (a,c) & 5\degree & $\approx$(a,c) ~dia & 0.842 & 222\\ 
 &  &  & $\approx$(a,c) ~dia & 0.842 & 222\\ 
 &  &  & $\approx$(a,c) ~dia & 0.800 & 218\\ 
 &  &  & $\approx$(a,c) dia & 0.800 & 218\\ \hline
a & (a,c) & 5\degree & $\approx$(-a,c) ~dia & 0.848 & 214\\ 
&  &  & $\approx$(-a,c) dia & 0.848 & 214\\ 
&  &  & $\approx$(-a,c) dia & 0.807 & 218\\ 
&  &  & $\approx$(-a,c) dia & 0.807 & 218\\ \hline
a & (a,b) & 3\degree & $\approx$ (-a,-b) dia & 0.962 & 130\\ 
&  &  & $\approx$(-a,-b) dia & 0.962 & 130\\
&  &  & $\approx$(-a,-b) dia & 0.964 & 130\\
&  &  & $\approx$(-a,-b) dia & 0.964 & 130\\ \hline
c & (a,c) & 3\degree & $\approx$(a,c) dia & 0.921 & 134\\ 
&  &  & $\approx$(a,c) dia & 0.921 & 134\\ 
&  &  & $\approx$(a,c) dia & 0.794 & 137\\ 
&  &  & $\approx$(a,c) dia & 0.794 & 137\\ \hline

a & (a,b) & 1\degree & $\approx$(a,b) ~dia & 0.569& 58\\ 
& & & $\approx$(a,b) dia & 0.569 & 58\\ 
 & & & $\approx$(-a,-b) dia & 0.572 & 58\\ 
&  &  & $\approx$(-a,-b) dia & 0.572 & 58\\ \hline
b & (a,b) & 4.25\degree & $\approx$(-a,-b) dia & 0.899 & 181\\ 
& &  & $\approx$(-a,-b) dia & 0.899 & 181\\ 
&  &  & $\approx$(-a,-b) dia & 0.900 & 181\\ 
&  &  & $\approx$(-a,-b) dia & 0.900 & 181\\ \hline

b & (b,c) & 4.25\degree & $\approx$(-b,c) ~dia & 0.809 & 188\\ 
 & &  & $\approx$(-b,c) ~dia & 0.809 & 188\\ 
&  &  & $\approx$(-b,c) ~dia & 0.872 & 183\\ 
 & &  &$\approx$ (-b,c) ~dia & 0.872 & 183\\ \hline

c & (b,c) & 4.25\degree & $\approx$(-b,c) ~dia & 0.802 & 188\\ 
& &  & $\approx$(-b,c) ~dia & 0.802 & 188\\ 
& &  & $\approx$(-b,c) ~dia & 0.864 & 186\\
 &  &  & $\approx$(-b,c) ~dia & 0.864 & 186\\ \hline
\end{tabular}
\caption{Static calculation EFG parameters for Model D using GGA+SOC+U. ``Axis'' labels the axis along which the oxygen atoms reside that is tilted and ``Plane'' labels the plane along which they are tilted. Angle $\phi$ is the tilt angle.} 
\label{tab:6}
\end{table}

\subparagraph{{\bf V.  Model E}}
In Model E, rotational distortions in the ($a$,$b$)-plane and tilt distortions in the ($a$,$c$)-plane are made. The point charge approximation\cite{Liu_PRB_2018} finds that a tilt angle of $\theta$ $\approx$ 8.5$\degree$ and a rotational angle of $\phi$ $\approx$ 12$\degree$ can produce EFG parameters that match NMR experiments. Nevertheless, our DFT+U calculations conflict with these earlier predictions. In fact, much as with Model D, our DFT+U simulations find that the principal axes for Model E also deviate from the crystalline coordinate axes and therefore disagree with experiments. This discrepancy may stem from difficulties converging the electric potential within a finite box in our original point charge approximation calculations. Representative data is presented in Table \ref{tab:7}.

\begin{table}[h]
\begin{tabular}{|c|c|c|c|c|}
\hline
\boldmath{$\theta$} & \boldmath{$\phi$} & \boldmath{$V_{zz}$} & \boldmath{$\eta$} &\boldmath{$\nu_Q \, {\rm (kHz)}$}           \\ \hline
5$\degree$ & 10$\degree$ & $\approx$(a,c) ~dia & 0.515 & 480 \\ 
 &  & $\approx$[$\bar{1}11$] dia & 0.897 & 621 \\ 
 & & $\approx$(-a,-c) dia & 0.570 & 483 \\ 
 & & $\approx$(-a,-c) dia & 0.562 & 484 \\ \hline

8.5$\degree$ & 12$\degree$ & $\approx$ (a,c) ~dia & 0.376 & 493 \\ 
& &$\approx$ (a,c) ~dia & 0.376 & 493 \\ 
 &  & $\approx$ (a,c) ~dia & 0.371 & 494 \\ 
 &  &$\approx$ (a,c) ~dia & 0.371 & 494 \\ \hline

7.8$\degree$ & 15$\degree$ & $\approx$ (a,c) ~dia & 0.249 & 547\\ 
&  & $\approx$ (a,c) ~dia & 0.249 & 547 \\ 
&  & $\approx$ (a,c) ~dia & 0.245 & 548 \\ 
 & & $\approx$ (a,c) ~dia & 0.245 & 548 \\ \hline
\end{tabular}
\caption{Static calculation EFG parameters for Model E using GGA+SOC+U. $\theta$ and $\phi$ denote the tilt and rotation angles.}
\label{tab:7}
\end{table}

\subparagraph{{\bf VI. Model F}}
Model F possesses a GdFeO$_3$-type distortion, which is common within perovskite oxides.\cite{glazer1972classification} We found that, again, the Model F principal axes align along diagonal directions, in disagreement with experimental observations. We also considered the flexible octahedra Model F2, which supplements the Model F GdFeO$_3$-type distortion with Model A-type elongations and compressions. We found that when the additional Model A type distortion is taken as -0.525\%, 0\%, and 0.52\% along the $a$, $b$, and $c$ axes respectively, which is the distortion that best matches experiments, Model F2 also produces experimentally-plausible EFG parameters. This is consistent with the notion that the main source of the observed non-zero EFG is from Model A distortions. 

\begin{table}[h]
\begin{tabular}{|c|c|c|c|}
\hline
 & \boldmath{$V_{zz}$} & \boldmath{$\eta$} &\boldmath{$\nu_Q \, {\rm (kHz)}$}           \\ \hline
F & $\approx$ (-a,-b) ~dia & 0.359 & 22 \\ 
& $\approx$ (-a,-b) ~dia & 0.349 & 23 \\ 
& $\approx$ [111] ~dia & 0.340 & 15 \\ 
& $\approx$ [111] ~dia & 0.346 & 15 \\ \hline
F2 & \,c & 0.954 & 186 \\
 & \,c & 0.954 & 186 \\
 &-a & 0.948 & 191 \\ 
 &-a& 0.948 & 191 \\ \hline
\end{tabular}
\caption{Static calculation EFG parameters for Models F and F2 using GGA+SOC+U. The data for Model F is shown for an $a^-a^-a^-$ of 5 degrees using Glazer's notation.\cite{glazer1972classification} Model F2 supplements Model F with Model A-type distortions.}
\label{tab:8}
\end{table}

To summarize, from static calculations of Models A-F2, we found that Model A, comprising a local distortion of the Na octahedra with Na-O bond elongation along the $a$ axis and compression along the $c$ axis of about 0.52\%, can best account for the EFG parameters obtained from NMR experiments. This orthorhombic distortion, which involves the three axes of the octahedra, a$\rightarrow$a+$\delta$, b$\rightarrow$b-$\delta$,  and  c$\rightarrow$c, corresponds to a static Q2 distortion mode \cite{khomskii2014transition}. 
In the presence of weak SOC, the Q2 and Q3 distortion modes lead to splitting of both  the  $t_{2g}$ and $e_{g}$ levels. While the former mode gives rise to orthorhombic local symmetry, the latter induces tetragonal local symmetry. Therefore, the Model A orthomhombic distortion corresponds to the Q2 distortion mode that splits the $t_{2g}$ levels into three singlets if the Jahn-Teller energy dominates \cite{khomskii2014transition}.  
The conjecture that Model A corresponds best to the  Q2 mode is supported by examining the physical origin of  the asymmetry parameter  $\eta \approx 1$.   
Since $\eta$ is defined as (V$_{xx}$-V$_{yy}$)/V$_{zz}$ and the sum of these three components must be zero, $\eta \approx 1$ implies that the smallest component of the EFG must be close to zero while the other two components must be equal in magnitude and opposite in direction, which, given that the principal axes of the EFG coincide with those of  the crystal,  intuitively leads to the Q2 mode described above. In systems without strong  spin orbit  coupling, such as $3d$ systems, the Jahn-Teller energy dominates over SOC. In this case,    the $d$ level will be split by the orthorhombic Q2 distortion  into three  singlets ($|d_{xy}\rangle$, $|d_{yz}\rangle$, and $| d_{xz}\rangle$). In systems with strong SOC, where the $d$ level   splitting is dominated  by  the spin orbit coupling, it is actually unclear if and how the degeneracy of the energy levels will be lifted by the  crystal field. 
The analysis of the magnetic entropy measurements in BNOO in \mbox{Ref. [\onlinecite{erickson2007FM}]} implied that the $j=\frac{3}{2}$ quartet is lifted to two Kramer doublets. Here, our EFG calculations reveal that there is a Q2 mode static distortion in the low temperature BLPS phase. This structural distortion   is concomitant  with the decrease of the degeneracy of the $j$ quartet, which leads us to suggest that this is a Jahn-Teller type structural distortion, in the sense that, instead of the orbital degeneracy in the weak SOC case, here, in the strong SOC case, it is a degeneracy of the total effective moment $j$ that is lifted  to reduce the total energy of the system. However,  we cannot provide proof of this hypothesis since a systematic theoretical framework for the description of the  spin orbit channels in the strong SOC case is lacking.

\subsection{EFG Tensor Predictions' Sensitivity to Distinct Magnetic Orders}
\label{subsec:cFM_magnetic_ordering}

In the following subsections, we will investigate the sensitivity of the EFG tensor  for the Model A.3 distortion to 
the presumed underlying magnetic structure.  

As described in the preceding sections, the experimental EFG parameters are overwhelmingly determined by the magnitude of the Jahn-Teller-type lattice distortion. In comparison, SOC and magnetic order are of secondary importance (see Table~\ref{tab:3}) as they only affect the linewidth, and not the splitting, in the leading order. However,  we tested the robustness of this conclusion to the assumed  magnetic    order. 

In general, the DFT+U method can produce multiple meta-stable solutions that reside in local energy minima.\cite{DFT+U_1,DFT+U_2} For this reason, when we polarize our initial magnetic moments along the cFM or FM[110] directions, we usually find that the final moments converge to approximately the same directions as the initial ones. The magnetic moments do not continuously change direction by much during the self-consistency cycle.
However, comparison of total energies of converged solutions, each with different moment directions, can reveal the polarizations' preferred easy-axes or planes (magnetic anisotropy energy or MAE). The MAE difference is expected to be on the order of tens of meV in the absence of SOC, or on the same order of the DFT precision. Therefore, the spins can rotate almost without energy cost in the absence of SOC.

To make contact with the results of Refs.~\onlinecite{Pickett_2015,Pickett_2016,Pickett_2007}, which found that the single sublattice FM110 was the energetically favored easy-plane using an onsite hybrid functional based on PBE, we performed static calculations with FM110 for the BLPS and undistorted cubic structures, both using GGA+SOC+U and PBE+SOC+U. Corresponding PBE+SOC+U results are given  in \mbox{Appendix \ref{FM110_PBE_A_22}}. 
The EFG tensor using PBE+SOC+U turns out to be smaller by a factor of a third compared with that produced using GGA+SOC+U. 

\begin{table}[h]
    \centering
    \begin{tabular}{|c|c|c|c|}
    \hline
    {\bf FM110} & \boldmath{$\nu_Q$}\bf{(kHz)} & \boldmath{$\eta$} & \boldmath{$V_{zz}$} \\ \hline \hline
    {\bf A} & 185 & 0.853 & a \\ \hline
    {\bf B Os1}  & 190 & 0.750 & a \\ \hline
    {\bf B Os2}  & -213 & 0.899 & a \\ \hline
    {\bf F2 Os}  & -210 & 0.893 & c \\ \hline
    {\bf Cubic Os} & 58 & 0.62 & [110] \\ \hline
    \end{tabular}
    \caption{Face-centered cubic FM110 EFG parameters from GGA+SOC+U. The similarity of these EFG tensors to those  for the cubic cell cFM in Tables~\ref{tab:2}, \ref{tab:4}, and \ref{tab:8} indicates that the EFG is insensitive to the precise nature of the magnetic order. }
    \label{tab:fcc_FM110_EFG_parameters}
\end{table}

As tabulated in Table~\ref{tab:fcc_FM110_EFG_parameters}, the EFG tensors for the representative BLPS models in the FM110 phase are similar to the cFM phase tensors, including the principal axes. Thus, 
as far as the EFG is concerned, the type of magnetic order in the presence of SOC does not lead to an appreciable modification of the charge density compared to that for the paramagnetic phase.

\subsection{EFG Tensor Predictions Using Different Density Functionals} 
\label{PBE0}

To check that our EFG parameters are indeed independent of the functional approximations employed, we present the EFG tensor obtained with the PBE0 hybrid functional for representative BLPS structures in the {\it fcc} primitive cell. The set magnetic order is FM001, and no SOC is considered, as calculations with the inclusion of SOC exceeded our computational resources. As presented in Table~\ref{tab:PBE0}, we find that Model A.3's splitting parameter, $\nu_Q$, is larger than that observed in experiments and the asymmetry factor $\eta$ is reduced to half of its measured  value.

To see if other distortion ratios are more consistent with experiment, we explored another four types of Model A distortions labelled as A.3.1 to A.3.4, representing elongation and compression along the $a$ and $c$ axes of 0.4\%, 0.6\%, 0.65\%, and 0.7\% (in absolute magnitude) of the Na-O bond length.  We see that as the distortion increases, both the splitting and asymmetry factors increase, which is in disagreement with the experimentally determined values. Model F2 using the PBE0 hybrid functional produces an $\eta$ value that matches experiment, yet a splitting much smaller than the measured value of \mbox{190-200 kHz}, which may merit further study. Regardless of the reduced asymmetry factor obtained from Model A type distortions, however, we always obtain a larger gap using PBE0. As a matter of fact,  using the GGA+SOC+U method, the gap obtained for Model A.3 with $U=3.3\,  {\rm eV}$ is 0.06 eV. 
Due to the neglect of SOC in our hybrid calculations, we cannot ascribe gap enhancement to the PBE0 functional alone. We additionally also calculated the EFG tensor of Model A.3 with the hybrid functional HSE06 and found that it gave the same EFG tensor as PBE0, while the gap was reduced to 0.3 eV.

\begin{table}[h]
    \centering
    \begin{tabular}{|c|c|c|c|c|}
    \hline
    {\bf fcc} & \boldmath{$\nu_Q$}\bf{(kHz)} & \boldmath{$\eta$} & \boldmath{$V_{zz}$} & {\bf gap (eV)} \\ \hline
    {\bf A.3} & 260 & 0.460 & -a & 1.0 \\ \hline
    {\bf A.3.1} & 233 & 0.396 & -a & 1.0 \\ \hline
    {\bf A.3.2} & 275 & 0.521 & -a & 1.28 \\ \hline
    {\bf A.3.3} & 293 & 0.594 &  \,a & 1.30  \\ \hline
    {\bf A.3.4} & 304 & 0.596 & -a & 1.30 \\ \hline
    {\bf F2}  & 138 & 0.882 &  \,c & 0.0 \\ \hline
    \end{tabular}
    \caption{EFG tensor for the {\it fcc} structures using the hybrid functional PBE0 for different distortions of Models A and F2.}    
    \label{tab:PBE0}
\end{table}

\section{Conclusion}
\label{Conc}

In this work, we carried out DFT+U calculations on the magnetic Mott insulator Ba$_2$NaOsO$_6$, which has strong spin orbit coupling. Our numerical work is inspired by recent NMR experiments on the material showing that it exhibited a broken local point symmetry  (BLPS) phase followed by an  exotic canted ferromagnetic order. Since earlier studies using the point charge approximation\cite{Liu_PRB_2018} were unable to distinguish between actual ion displacement and charge density deformations, the nature of the BLPS phase remained unclear. In this paper, with the input of EFG parameters obtained from NMR experiments, we were able to explicitly show that the main source of the non-zero EFG parameters observed in NMR experiments is an orthorhombic local distortion (corresponding to a Q2 distortion mode) of the Na-O octahedra.   {This distortion  is insensitive to the type of underlying magnetic order and } lifts the $j = \frac{3}{2}$ quartet to two Kramer doublets before the onset of cFM order. We attribute this distortion to a kind of Jahn-Teller effect which, instead of lifting the orbital degeneracy in the weak SOC case, lifts the degeneracy of the total effective moment $j$ to Kramer doublets, since the magnitude of the spin orbit coupling energy is, in this case, larger than the magnitude of the Jahn-Teller energy.

Moving forward, it would be worthwhile to more thoroughly investigate the cFM order observed in this work using other functionals or methods more adept at handling strong correlation to eliminate any ambiguities that stem from our specific computational treatment.  Another future direction would be the {\it ab-initio} calculation of the NMR hyperfine tensor, which captures the electron-nuclear spin-spin interaction at the sodium site.  We expect that the understanding of the interplay between spin, orbital, and lattice degrees of freedom in the 5$d^1$ magnetic Mott insulator forged in this work will be of crucial importance for understanding experiments on related transition metal compounds, such as Ba$_2$LiOsO$_6$ and BaCaOsO$_6$, already underway.

  \section{Acknowledgments}
\label{Ack}

The authors thank Jeong-Pil Song and Yiou Zhang for enlightening discussions.  We are especially grateful to Ian Fisher for the long term collaboration on the physics of \BaOs. This work was supported in part by U.S. National Science Foundation grants DMR-1608760 (V.F.M.) and DMR-1726213 (B.M.R.). The calculations presented here were performed using resources at the Center for Computation and Visualization, Brown University, which is supported by NSF Grant No. ACI-1548562.

  \section{Appendix}
      \subsection{\BaOs Unit Cell}
      \label{Unitcell}
        \vspace*{-0.2cm}

\begin{center}	 
\begin{figure}[b]
  \vspace*{-0.0cm}
\begin{minipage}{0.98\hsize}
 \centerline{\includegraphics[scale=0.50]{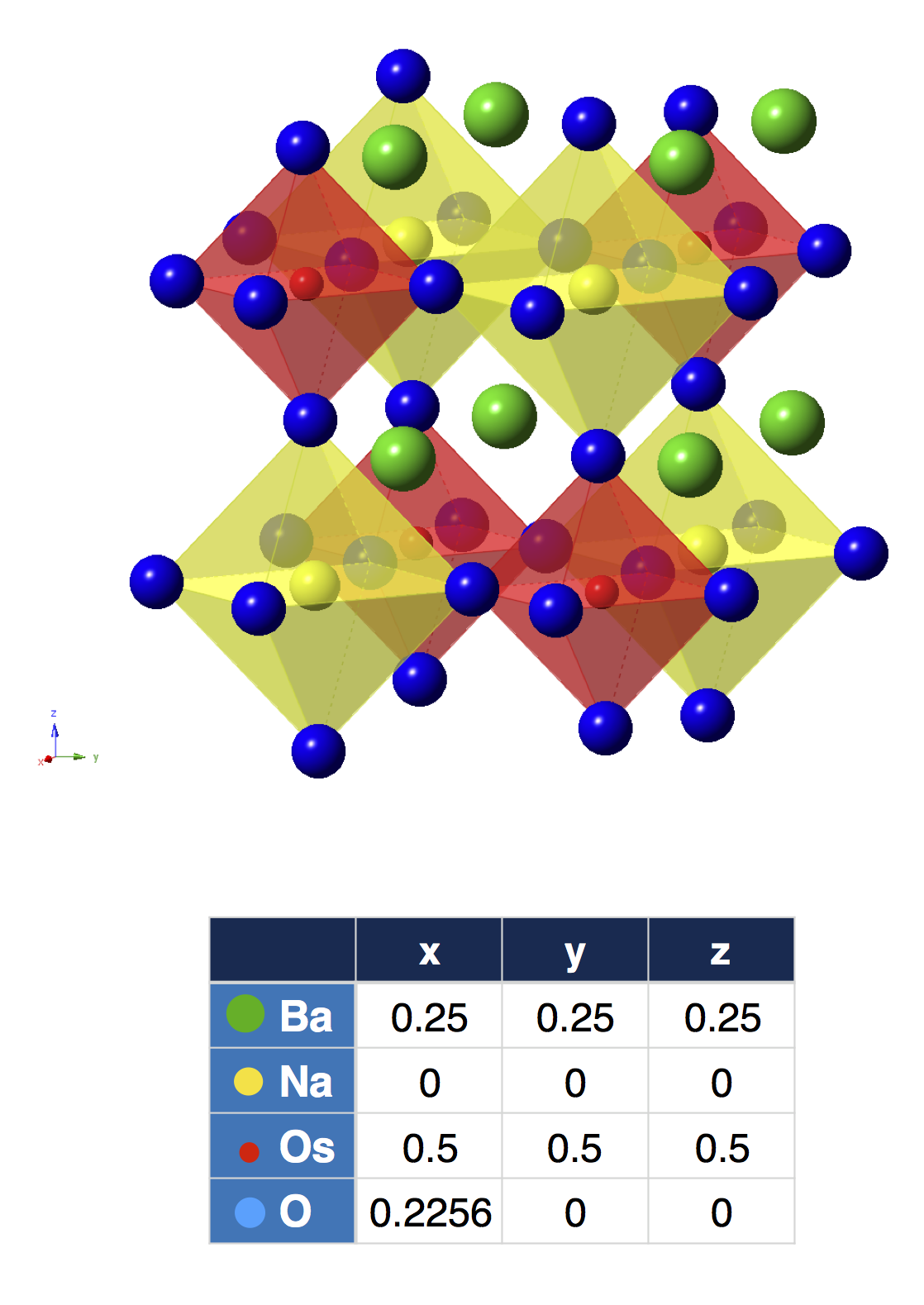}} 
\begin{minipage}{.98\hsize}
 \vspace*{-0.0cm}
\caption[]{\label{unitcell} \small 
Unit cell of Ba$_2$NaOsO$_6$, consisting of 40 total atoms. 
Barium atoms are indicated by green, sodium by yellow, osmium by red, and oxygen by blue. Osmium and sodium octahedra are shaded in red and yellow, respectively. The position of the atoms in the unit cell is shown in the bottom panel. }
 \vspace*{-0.0cm}
\end{minipage}
\end{minipage}
\end{figure}
%
\end{center}

The crystal structure of \BaOs, as deduced from X-ray diffraction at room temperature
 \cite{erickson2007FM,stitzer2002crystal}, is shown in \mbox{Fig. \ref{unitcell}}.  
At room temperature, this material possesses an undistorted double-perovskite structure, in which the OsO$_6$ octahedra are neither distorted nor rotated with respect to each other or the underlying lattice.   This  undistorted double-perovskite structure belongs to the $Fm\bar3 m$ space group. 
In our calculations,   a simple cubic cell consisting of four $fcc$ primitive cells is modeled so as to accommodate the emergence of a cFM phase and its related distortions.   As a matter of fact, if we consider only one  type of distortion, as is the case for all models except Model B, a simple cubic cell consisting of two $fcc$ primitive cells suffices to correctly compute the EFG parameters. For Model B, which comprises two types of distortion, four $fcc$ primitive cells are required.

  \subsection{Check on DFT+U Parameterization}
  \label{CheckDFT}
  
  We summarize the calculated variation of EFG parameters $\nu_Q$, $\eta$, and gap with $U$ and $J$ for Model A.3 with GGA+SOC+U using the PP6 pseudopotential. We find that within  this formalism, $\eta$ shows a larger variation with changes in $U$ and $J$ than $\nu_Q$.  
  
\vspace*{-0.1cm}
\begin{table}[h!]
\centering
\begin{tabular}{|c|c|c|c|c|}
\hline
 \bf{U (eV)} & \bf{J (eV)} & \boldmath{$\nu_Q$} \bf{(kHz)} & \boldmath{$\eta$} & \bf{gap (eV)} \\ \hline 
 3.3 & 0.5 & 194 & 0.866 & 0.06 \\ \hline 
 4.0 & 0.5 & 194 & 0.873 & 0.244 \\ \hline 
 4.5 & 0.5 & 193 & 0.863 & 0.388 \\ \hline
 5.0 & 0.5 & 190 & 0.852 & 0.556 \\ \hline
 3.3 & 0.6 & 191 & 0.819 & 0.04 \\ \hline
\end{tabular}
\caption{The variation of EFG parameters $\nu_Q$, $\eta$ and gap with $U$ and $J$ for Model A.3 with GGA+SOC+U  using the  PP6 pseudopotential.  }
\label{SItab:EFG_vary_U_J}
\end{table}

 \vspace*{-0.4cm}
\subsection{DFT Convergence Tests}
\label{ConvTest}

Convergence tests were performed for Model A with elongation of the Na-O bond along the $x,y,z$ axes of 0.55\%, 0.1\%, and 0.2\%. The projector augmented wave (PAW) pseudopotential used was Ba$_{sv}$+Na+Os+O  with  \mbox{$U = 3.3$ eV} and \mbox{$J=0.5$ eV}. The cutoff energy for the planewave basis set  was  \mbox{600  eV} and the global break condition for the electronic self-consistency loop was  \mbox{$10^{-5}$ eV}. We found that the energy converges with an 8$\times$8$\times$8 k-point grid or finer.

\subsection{Static Calculations of Models A.3 and B.2  Using PBE+SOC+U, FM [110]}
\label{StatCalcs}
\label{FM110_PBE_A_22}

We summarize the calculated variation of EFG parameters   for distortion Models A.3 and B2 using PBE+SOC+U 
in the one-sublattice FM[110] state.

 \noindent {\it Model A.3, PBE+SOC+U, FM [110] }


\begin{table}[!h]
\centering
\begin{adjustbox}{max width=\columnwidth}
\begin{tabular}{|c|c|c|c|c|c|c|}
\hline
\bf{Atoms} & \boldmath{$V_{xx}$} & \boldmath{$V_{yy}$} &\boldmath{ $V_{zz}$} & \boldmath{$\nu_Q$} \bf{(kHz)} & \boldmath{$\eta$} & \boldmath{$V_{zz}$} \bf{axis}  \\ \hline
\bf{Na1} & -0.799 & -0.087 & 0.886 & 128 & 0.803 & a \\ \hline
\bf{Na3} & -0.799 & -0.087 & 0.886 & 128 & 0.803 & a \\ \hline
\end{tabular}
\end{adjustbox}
\caption{EFG parameters (V$_{xx}$, V$_{yy}$, and V$_{zz}$ are in units of V/A$^2$) calculated  using  PBE+SOC+U (Gap = 0.33 eV)  for Model A.3.}
\label{table_A_22_FM110_PBE_EFG}
\end{table}


\noindent{\it Model B.2, PBE+SOC+U, FM [110]}

 \vspace*{-0.4cm}

\begin{table}[!h]
\centering
\begin{adjustbox}{max width=\columnwidth}
\begin{tabular}{|c|c|c|c|c|c|c|}
\hline
\bf{Atoms} & \boldmath{$V_{xx}$} & \boldmath{$V_{yy}$} &\boldmath{ $V_{zz}$} & \boldmath{$\nu_Q$} \bf{(kHz)} & \boldmath{$\eta$} & \boldmath{$V_{zz}$} \bf{axis} \\ \hline
\bf{Na1} & -0.753 & -0.144 & 0.896 & 130 & 0.679 & a  \\ \hline
\bf{Na3} & 0.962 & 0.063 & -1.025 & 150 & 0.877 & a \\ \hline
\end{tabular}
\end{adjustbox}
\caption{EFG parameters (V$_{xx}$, V$_{yy}$, and V$_{zz}$ are in units of V/A$^2$) calculated  using PBE+SOC+U (Gap = 0.31 eV)  for Model B.3.}
\label{table_B_FM110_PBE_EFG}
\end{table}

\vspace*{-0.0cm}
\subsection{Summary of BNOO Optimization Calculations}
\label{Optimization}
\vspace*{-0.3cm}
In addition to calculating these parameters for the static distorted structures described in Section~\ref{Results}, we have also calculated the EFG parameters after relaxing these structures' geometries. The $\eta$ and $\nu_Q$ are reduced in general. In the case of Model A.3, the energy of the relaxed structure is lowered by \mbox{0.3 eV}, while the EFG parameters are reduced to roughly half of their unrelaxed values. This is because the relaxation tends to reduce the size of distortion. \\

 $^\dag$ Corresponding authors V. F. M (vemi@brown.edu)  \&  B. R. (brenda\underline{ }rubenstein@brown.edu)
 
\bibliography{ref}



\end{document}